\documentclass[aps,prx,preprint,onecolumn,citeautoscript,superscriptaddress,footinbib,
eqsecnum]{revtex4-1}  
\usepackage{amsmath,amssymb,bm,bbm} 
\usepackage{amsthm}
\usepackage{graphicx}  
\usepackage{color} 
\usepackage{bbm}
\usepackage[dvipsnames]{xcolor}
\usepackage[papersize={8.5in,11in}]{geometry}
\usepackage[colorlinks=true]{hyperref}  
\usepackage[section]{placeins}
\hypersetup{
    bookmarks=true,         % show bookmarks bar?
    unicode=false,          % non-Latin characters 
    pdftoolbar=true,        % show Acrobat
    pdfmenubar=true,        % show Acrobat 
    pdffitwindow=false,     % window fit to page when opened
    pdfstartview={FitH},    % fits the width of the page to the window
    pdftitle={Z2 fractionalized phases of a solvable, disordered, $t$-$J$ model},    % title
    pdfauthor={Wenbo Fu, Yingfei Gu, Subir Sachdev, Grigory Tarnopolsky},     % author
    pdfsubject={},   % subject of the document
    pdfcreator={},   % creator of the document
    pdfproducer={}, % producer of the document
    pdfkeywords={} {} {}, % list of keywords
    pdfnewwindow=true,      % links in new window
    colorlinks=true,       % false: boxed links; true: colored links
    linkcolor=magenta, %red,          % color of internal links (change box color with linkbordercolor)
    citecolor=blue,        % color of links to bibliography
    filecolor=magenta,      % color of file links
    urlcolor=blue           % color of external links
} 

\usepackage{tikz}
\usepackage{tikz-cd}
\usetikzlibrary{arrows}
\usetikzlibrary{intersections}
\usetikzlibrary{shapes.geometric}
\usetikzlibrary{decorations.pathmorphing, patterns,shapes}

\renewcommand{\bar}{\overline}

\renewcommand{\leq}{\leqslant}

\newcommand{\nn}{\nonumber\\}

\newcommand{\be}{\begin{equation}}
\newcommand{\ee}{\end{equation}}
\newcommand{\bea}{\begin{eqnarray}}
\newcommand{\eea}{\end{eqnarray}}

\def\veps{\varepsilon}
\def\nn{\nonumber\\}
\newcommand{\R}{{\mathrm{R}}}
\newcommand{\NS}{{\mathrm{NS}}}
\def\fr#1{(\ref{#1})}
\def\c{{\bf C}}

\usepackage{amsfonts}
\usepackage{upgreek}
\usepackage{slashed}
\usepackage{latexsym}
\usepackage{todonotes}

\newcommand{\beq}{\begin{equation}}
\newcommand{\eeq}{\end{equation}}
\def\bea{\begin{eqnarray}}
\def\eea{\end{eqnarray}}

\begin{document}

\title{NMR relaxation in Ising spin chains}
\author{Julia Steinberg}
\affiliation{Department of Physics, Harvard University, Cambridge MA 02138, USA}

\author{N. P. Armitage}
\affiliation{The Institute for Quantum Matter, Department of Physics and Astronomy,
The Johns Hopkins University, Baltimore, MD 21218 USA}

\author{Fabian H.L. Essler}
\affiliation{Rudolf Peierls Centre for Theoretical Physics, Parks Road, Oxford OX1 3PU, UK}

\author{Subir Sachdev}
\affiliation{Department of Physics, Harvard University, Cambridge MA 02138, USA}
\affiliation{Perimeter Institute for Theoretical Physics, Waterloo, Ontario, Canada N2L 2Y5}

\date{\today
\\
\vspace{0.4in}}

\begin{abstract}
We examine the low frequency spin susceptibility of the paramagnetic phase of the quantum Ising chain in transverse field at temperatures well below the energy gap. We find that the imaginary part is dominated by rare quantum processes in which the number of quasiparticles changes by an odd number. We obtain exact results for the NMR relaxation rate in the low temperature limit for the integrable model with nearest-neighbor Ising interactions, and derive exact universal scaling results applicable to generic Ising chains near the quantum critical point.   These results resolve certain discrepancies between the energy scales measured with different experimental probes in the quantum disordered paramagnetic phase of the Ising chain system CoNb$_2$O$_6$.
\end{abstract}

\maketitle

\section{Introduction}
\label{sec:intro}

The transverse field Ising chain is an ideal setting to study the dynamics of quantum criticality \cite{ssbook} as many observable properties can be computed either exactly, or reliably in a semiclassical approach \cite{SSAPY97,Essler09,Iorgov10}.  In recent years, there have been several experimental realizations of the transverse Ising chain that make theoretical predictions testable
\cite{Coldea10,bernien2017probing}.  For instance, it has been found that CoNb$_2$O$_6$ is for many purposes an almost ideal realization of the one-dimensional ferromagnetic Ising chain.  Experiments have studied its properties across the different regimes of the phase diagram as a function of transverse field and temperature \cite{Coldea10,Cabrera14,Imai14,morris2014hierarchy,liang2015heat,robinson2014quasiparticle}.

\begin{figure}
\begin{center}
\includegraphics[width=5.5in]{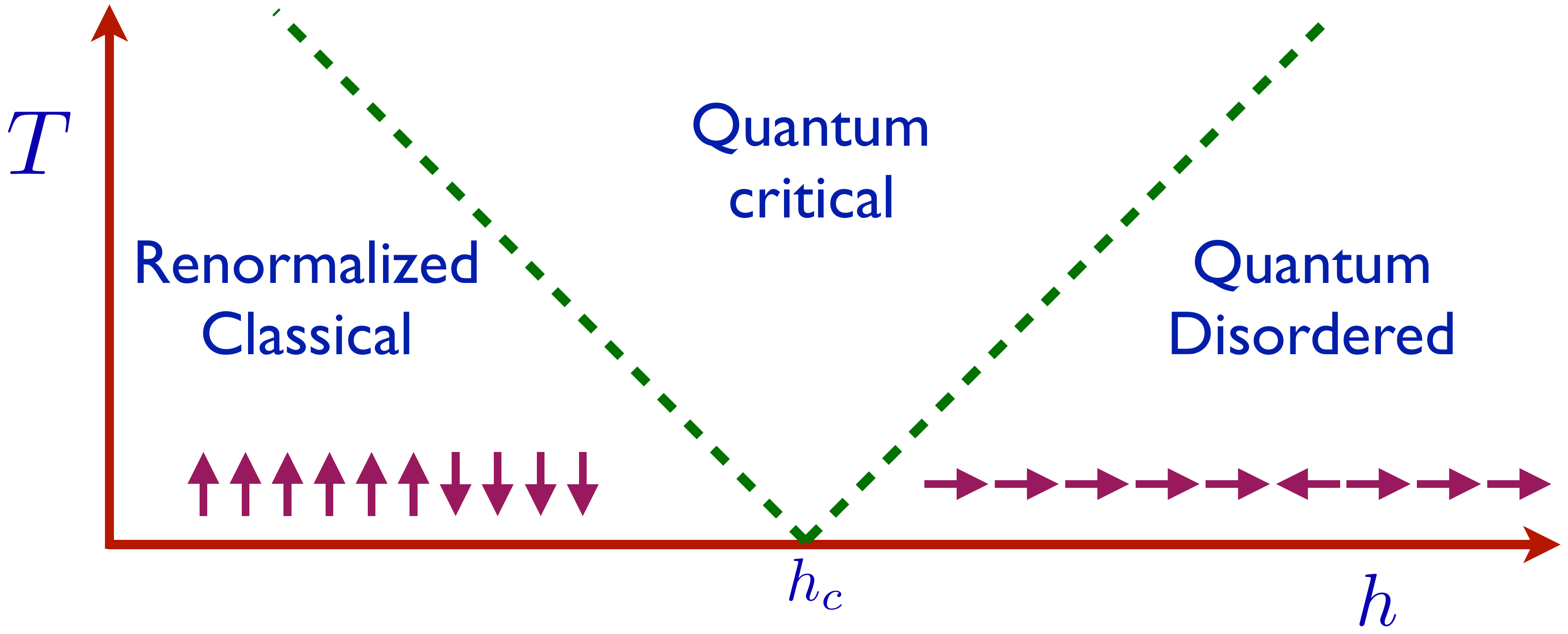}
\end{center}
\caption{Crossover phase diagram of the Ising chain in a transverse field, $h$. There is a quantum critical point at $T=0$ and $h=h_c$ (for the Hamiltonian in
Eq.~(\ref{HI}), $h_c=1$) between a ferromagnetic ($h<h_c$) which breaks the Ising symmetry, and a paramagnetic phase. This paper focuses on the ``quantum disordered'' regime above the paramgetic phase for $h>h_c$. The other regimes were described in Ref.~\onlinecite{Imai14}.} 
\label{fig:ising}
\end{figure}

In this paper, we revisit the issue of the nuclear magnetic resonance (NMR) 
relaxation in an Ising spin chain in its gapped state without ferromagnetic order (the `quantum disordered'
regime of Fig.~\ref{fig:ising}). A recent NMR experiment on CoNb$_2$O$_6$ \cite{Imai14} studied the NMR relaxation rate, $1/T_1$, in all three regimes near the quantum critical point of the phase diagram in Fig.~\ref{fig:ising}. Experimental results agreed quantitatively with theoretical predictions in the `renormalized classical' and `quantum critical' regimes. While there were no firm theoretical predictions in the quantum disordered regime, it was conjectured \cite{Imai14} that the low temperature ($T$) behavior was $1/T_1 \sim \exp(-\Delta/T)$, where $\Delta$ is the energy gap to single spin flips. However, other experiments probing the large field transverse paramagnetic regime show discrepancies with the energy scales probed by NMR. The NMR experiments \cite{Imai14} measured an activation energy that was approximately two times larger than the gap inferred from heat capacity \cite{liang2015heat}, neutron scattering \cite{robinson2014quasiparticle}, and THz/infrared experiments \cite{Virok2018}.

Here we will show that near the critical point the behavior of the NMR relaxation rate is in fact $1/T_1 \sim \exp(-2\Delta/T)$, and compute the precise prefactor for the integrable nearest-neighbor Hamiltonian. We find that the result is compatible with the universal
relativistic quantum field theory, and obtain the universal behavior of $1/T_1$ at $T \ll \Delta$ for a generic Ising Hamiltonian. 
%Add a bit about resolving issue by shedding light on dominant processes
We begin by recalling some exact results on the nearest-neighbor Ising chain in Section~\ref{sec:exact}. In particular, the lattice form factors computed in Ref.~\onlinecite{Iorgov10} will be crucial ingredients in our results.
The computation of the NMR relaxation rate of the nearest-neighbor Ising chain appears in Section~\ref{sec:nmr}. Section~\ref{sec:qc} describes the universal behavior of the NMR relaxation rate across the quantum critical point.  The experimental situation is discussed in Section~\ref{sec:exp}.

\section{Exact spectrum}
\label{sec:exact}

We work with the integrable Ising chain Hamiltonian
\beq
H = - J \sum_{\ell=1}^L \left[ \sigma^z_\ell \sigma^z_{\ell+1} + h \sigma^x_\ell \right] \label{HI}
\eeq
where $\sigma^{x,z}_\ell$ are Pauli matrices acting on the 2-state spins on site $\ell$. For $h<1$, this model has a ferromagnetic ground state with $\langle \sigma^z_\ell \rangle = N_0 \neq 0$. We are interested in the low $T$ behavior in the paramagnetic state for $h>1$, where $\langle \sigma^z_\ell \rangle = 0$ at $T=0$.

The spectrum of $H$ can be computed exactly by a Jordan-Wigner transformation which maps it onto a theory of spinless fermions with dispersion
\beq
\varepsilon_p = 2 J \sqrt{ 1 + h^2 - 2 h \cos (p) }
\label{eq:disp}
\eeq
as a function of crystal momentum $-\pi < p < \pi$. This dispersion implies an energy gap
\beq
\Delta = 2J |h -1 | \,.
\eeq
The complete set of excited states are described by $n$ fermion states $\left|p_1, p_2, \ldots , p_n \right\rangle$, where all fermion momenta must be unequal.

Remarkably, all matrix elements of the ferromagnetic order parameter, $\sigma^z_\ell$, between all many-body states
have been computed exactly by Iorgov {\it et al.} \cite{Iorgov10}. For our purposes, we need the matrix elements in the limit $L \rightarrow \infty$, which can be written as
\bea
&& \left\langle q_1, \ldots , q_{2n} \right| \sigma^z_\ell \left| p_1 , \ldots , p_m \right\rangle = \frac{(4 J^2 h)^{(m-2n)^2 /4}}{L^{n+m/2}} |1 - h^2|^{1/8} i^{\lfloor n+ m/2 \rfloor} e^{- i \ell \left[ \sum_{j=1}^{2n} - \sum_{l = 1}^m p_l \right]} \nn
&& ~~~~~\times \prod_{j=1}^{2n} \frac{1}{\sqrt{\varepsilon_{q_j}}} \prod_{l=1}^m \frac{1}{\sqrt{\varepsilon_{p_l}}}
\prod_{j<j'=1}^{2n} \frac{2 \sin(q_j-q_{j'})}{\varepsilon_{q_j} + \varepsilon_{q_{j'}}} 
\prod_{l<l'=1}^{m} \frac{2 \sin(p_l-p_{l'})}{\varepsilon_{p_l} + \varepsilon_{p_{l'}}} 
\prod_{j=1}^{2n}\prod_{l=1}^{m} \frac{\varepsilon_{q_j} + \varepsilon_{p_{l}}}{2 \sin(q_j-p_{l})}\,,
\label{ff}
\eea
where $m$ is even (odd) for $h<1$ ($h>1$). We will be able to compute the NMR relaxation rate in the $T \rightarrow 0$ limit 
for $h>1$ by a direct application of Eq.~(\ref{ff}) in the Lehmann spectral representation.

\section{NMR relaxation rate}
\label{sec:nmr}

The NMR relaxation rate is determined by the low frequency behavior of local spin susceptibility.
We define the imaginary time ($\tau$) susceptibility by
\beq
\chi (\tau) = \int_0^{1/T} d \tau \left\langle \sigma^z_0 (\tau) \sigma^z_0 (0) \right\rangle \,. \label{chit}
\eeq
After a Fourier transform and analytic continuation to real frequencies ($\omega$), we obtain the NMR relaxation rate from \cite{Imai14}
\beq
\frac{1}{T_1} = \lim_{\omega \rightarrow 0} \frac{2 T}{ \omega} |a_{hf}|^2 \mbox{Im} \chi (\omega) \,, \label{nmr}
\eeq
where $a_{hf}$ is the hyperfine coupling between the nuclei and the Ising spins.

\subsection{Low-temperature expansion}
\label{ssec:lowT}
We are interested in the retarded two-point order parameter
autocorrelator 
\be
\chi(\omega)=\int_0^{1/T} d\tau e^{i\bar\omega\tau}\frac{1}{{\cal
    Z}}{\rm Tr}\left[ e^{- H/T}\sigma^z_{j}(\tau)\sigma^z_{j}\right]\Bigg|_{\bar\omega\rightarrow\eta-i\omega},
\ee
where ${\cal Z}$ is the partition function. The idea of Ref.~\onlinecite{Essler09} is
to develop a linked cluster expansion for this quantity. The starting
point is the Lehmann representation 
\be
\chi(\omega)=\frac{1}{{\cal Z}}\sum_{n,m=0}^\infty
C_{n,m}(\omega)\ ,
\label{chiSRfvol}
\ee
where 
\begin{eqnarray}
\label{eix}
C_{n,m}(\omega)&=& \frac{1}{n!}\sum_{\{k_1,\dots k_n\}}\frac{1}{m!}\sum_{\{p_1,\dots,p_m\}}
|\langle k_1\cdots k_{n}|\sigma^z_0|p_1\cdots p_{m}\rangle|^2
%\nn&&\times\ 
\frac{e^{-E(\{p_i\})/T}-e^{-E(\{k_j\})/T}}{\omega+i\eta-E(\{p_i\})+E(\{k_j\})}\ .
\end{eqnarray}
Here $\eta$ is a positive infinitesimal. The expansion of the partition function reads 
\be
{\cal Z}=1+\sum_{p\in \R}e^{-\veps_p/T}+
\sum_{p_1 < p_2\in
  \NS}e^{-[\veps_{p_1}+\veps_{p_2}]/T}+\ldots\equiv 
\sum_{n=0}^\infty{\cal Z}_n.
\ee
By construction the contribution ${\cal Z}_n$ scales with system size as $L^n$.
Here the subscripts refer to Ramond and Neveu-Schwartz boundary conditions, which will not matter in the infinite $L$ limit we take.
The individual terms $C_{n,m}(\omega)$ in the expansion \fr{chiSRfvol}
diverge with the system size $L$ because the matrix elements
\fr{ff} become singular when $k_r\rightarrow p_s$. Ref.~\onlinecite{Essler09} re-casts the expansion in terms of
\emph{linked clusters}, which are finite in the thermodynamic
limit. The linked clusters relevant for a low-temperature expansion of
$\chi(\omega)$ are
\bea
{\bf C}_{2n+1,0}(\omega)&=&C_{2n+1,0}(\omega)\ ,\qquad
{\bf C}_{0,2n+1}(\omega)=C_{0,2n+1}(\omega)\ ,\nn
{\bf C}_{1,2n}(\omega)&=&C_{1,2n}(\omega)-{\cal  Z}_1C_{0,2n-1}(\omega)\ ,\nn
{\bf C}_{2,2n+1}(\omega)&=&C_{2,2n+1}(\omega)-{\cal
  Z}_1C_{1,2n}(\omega)-({\cal  Z}_2-{\cal Z}_1^2)C_{0,2n-1}(\omega)\ ,\nn
{\bf C}_{3,2n}(\omega)&=&C_{3,2n}(\omega)-{\cal
  Z}_1C_{2,2n-1}(\omega)-({\cal  Z}_2-{\cal Z}_1)^2C_{1,2n-2}(\omega)\ .
\eea
In terms of the linked clusters we have
\be
{\rm Im}\ \chi(\omega)=\sum_{n,m=0}^\infty {\rm Im}\
   {\bf C}_{n,m}(\omega)\ .
\ee
The leading terms at low temperatures and $\omega\approx 0$ are
\bea
{\cal C}_1(\omega)&=&\c_{1,2}(\omega)+\c_{2,1}(\omega)\ ,\nn
{\cal C}_2(\omega)&=&\c_{2,3}(\omega)+\c_{3,2}(\omega)\ ,\nn
{\cal C}_3(\omega)&=&\c_{1,4}(\omega)+\c_{4,1}(\omega)+\c_{3,4}(\omega)+\c_{4,3}(\omega)\ .
\eea
As $\varepsilon_k>\Delta=2J|h-1|$ the formal temperature dependence of these
terms at $T\ll\Delta$ is
\be
{\cal C}_n(\omega)={\cal O}\big(e^{-(n+1)\Delta/T}\big).
\ee
As we are interested in the NMR relaxation rate we focus on the quantities
\be
c_n(T)=\lim_{\omega\to 0}\frac{ {\rm Im}\big({\cal
    C}_n(\omega)\big)}{\omega}\ .
\ee

%%%%%%%%%%%%%%%%%%%%%%%%%%%%%%%%%%%%%%%%%%%%%%%
\subsubsection{Leading term}
\label{sec:leading}
%%%%%%%%%%%%%%%%%%%%%%%%%%%%%%%%%%%%%%%%%%%%%%%
We begin with some qualitative considerations on the physical processes which lead to the dominant contributions to Eq.~(\ref{nmr}) as $T \rightarrow 0$ for $h>1$. A thermal excitation with energy $E_i$ will appear with a probability $e^{-E_i/T}$ as an initial state in the relaxation process. We should focus on the states with the lowest possible $E_i$.
Because of the $\omega \rightarrow 0$ limit, the final states will also have an energy $E_f = E_i$ and are reached by the action of the $\sigma^z_0$ operator. We notice from Eq.~(\ref{ff}) that for $h>1$ the matrix element is non-zero only between states with distinct parities in the number of fermions. Therefore the initial and final states must differ by an odd number of fermions which places strong constraints on the ranges of allowed values of $E_i$ and $E_f$.

We first consider the process $1 \rightarrow 2$, from an initial state with one fermion to a final state with 2 fermions.
This process (and its inverse) will dominate as $T \rightarrow 0$. The single fermion excitations are in the energy range $(\varepsilon_{\rm min}, \varepsilon_{\rm max}) \equiv 2 J (h-1, h+1)$. In the most optimal conditions, both fermions in the 2 particle state will have energy close to $\varepsilon_{\rm min}$. So for a $1 \rightarrow 2$ particle process to be allowed, we need $\varepsilon_{\rm max} > 2 \varepsilon_{\rm min}$ or $h< 3$. 
This process will have probability $\exp(-2 \varepsilon_{\rm min}/T)$.

For $h>3$ we need to consider processes with larger numbers of fermions to obtain the leading contribution. In general, the 
$n \rightarrow m$ process is allowed for $h< (m+n)(m-n)$, where $m>n$ and $m-n$ is odd. Such a process occurs with probability $\exp(-m \varepsilon_{\rm min}/T)$.
Thus for $3 < h < 5$ the most probable process is $2 \rightarrow 3$ with probability $\exp(-3 \varepsilon_{\rm min}/T)$.
There are also processes at smaller $h$, such as $1 \rightarrow 4$ for $h< 5/3$ with probability $\exp(-4 \varepsilon_{\rm min}/T)$.

We now consider the $1 \rightarrow 2$ process which has a prefactor of $\exp ( -2 \Delta/T)$. Importantly for $\omega\approx 0$ and in the limit $\eta\to 0$ we have
\be
\lim_{\eta\to 0}\big[{\rm Im}{\cal C}_1(\omega)\big]=\lim_{\eta\to 0} {\rm Im}\big[C_{1,2}(\omega)+C_{2,1}(\omega)\big]\,
\ee
i.e. the ``disconnected'' contributions ${\cal Z}_1 C_{01}(\omega)$ and ${\cal Z}_1 C_{10}(\omega)$ vanish in the limit $\eta\to 0$. This is related to the fact that the kinematic poles in the form factors do not contribute to the momentum sums by virtue of the energy-conservation delta function.
Using the explicit expression for the form factors \fr{ff} and turning momentum sums into integrals in the $L\to\infty$ limit we have
\bea
{\rm Im}\big({\cal C}_1(\omega)\big)&=&\frac{J\pi}{4}[h^2(h^2-1)]^\frac{1}{4}
    \int_{-\pi}^\pi
\frac{dp_1dp_2dq}{(2\pi)^3}
\left[\delta(\omega+\varepsilon_{p_1}+\varepsilon_{p_2}-\varepsilon_{q})
  -\delta(\omega-\varepsilon_{p_1}-\varepsilon_{p_2}+\varepsilon_{q})\right]\nn
&&\times\ \frac{\displaystyle (\varepsilon_{p_1}+\varepsilon_q)^2(\varepsilon_{p_2}+\varepsilon_q)^2\sin^2\left(\frac{p_1-p_2}{2}\right)}{ \displaystyle \varepsilon_{p_1}\varepsilon_{p_2}\varepsilon_{q}(\varepsilon_{p_1}+\varepsilon_{p_2})^2
  \sin^2\left(\frac{p_1-q}{2}\right)  \sin^2\left(\frac{p_2-q}{2}\right)}
\left[e^{-(\varepsilon_{p_1}+\varepsilon_{p_2})/T}-e^{-
    \varepsilon_q/T}\right].
\eea
Carrying out one of the integrals using the energy conservation delta-function we obtain
\bea
c_1(T)&=&\frac{J}{2T}[h^2(h^2-1)]^\frac{1}{4}
    \int_{-\pi}^\pi
    \frac{dp_1dp_2}{(2\pi)^2} \Theta_H\big(2J(h+1)-\varepsilon_{p_1}-\varepsilon_{p_2}\big)
    \frac{e^{-(\varepsilon_{p_1}+\varepsilon_{p_2})/T}}{|\varepsilon'_{p_0}|}\nn
&&\qquad\qquad\qquad\times\    \frac{\displaystyle (2\varepsilon_{p_1}+\varepsilon_{p_2})^2(\varepsilon_{p_1}+2\varepsilon_{p_2})^2\sin^2\left(\frac{p_1-p_2}{2}\right)}{\displaystyle \varepsilon_{p_1}\varepsilon_{p_2}(\varepsilon_{p_1}+\varepsilon_{p_2})^3
      \sin^2\left(\frac{p_1-p_0}{2}\right)
      \sin^2\left(\frac{p_2-p_0}{2}\right)}\ ,
\label{c1T}
\eea
where
\be
p_0={\rm arccos}\left[\frac{ \displaystyle 1+h^2-\left(\frac{ \varepsilon_{p_1}+\varepsilon_{p_2}}{2J}\right)^2}{2h}\right].
\ee
In the low-temperature limit $T\ll 2J(h-1)$ and $h<3$ we can carry out the
remaining two integrals as follows. The integration will be dominated by small $p_{1,2}\sim \sqrt{T}$. For these small $p$, we can expand the dispersion as
\beq
\varepsilon_p = \Delta + \frac{p^2}{2 m} + \ldots 
\eeq
where $\Delta=2J(h-1)$ and $m=(h-1)/(2hJ)$. Expanding the rest of the integrand around $p_1=p_2=0$ we obtain
\bea
c_1(T)&=&\frac{J}{T}e^{-2\Delta/T}\frac{81[h^2(h^2-1)]^\frac{1}{4}}{64|\varepsilon'_{q_0}|\Delta\sin^4({q_0}/{2})}
    \int_{-\pi}^\pi
    \frac{dp_1dp_2}{(2\pi)^2} e^{-\frac{p_1^2+p_2^2}{2mT}}(p_1-p_2)^2\ ,\nn
    &=&\frac{81m^2TJ[h^2(h^2-1)]^\frac{1}{4}}{64\pi\Delta\sin^4(q_0/2)|\varepsilon'_{q_0}|}e^{-2\Delta/T}\ ,
\eea
where we have defined
\be
q_0={\rm arccos}\left[\frac{1+h^2-4(h-1)^2}{2h}\right].
\ee
Using  Eq.~(\ref{nmr}), we obtain our main result
\beq
\frac{1}{T_1} = |a_{hf}|^2 \left[ \frac{81 J m^2 [h^2(h^2-1)]^{1/4}}{32 \pi \Delta |v| \sin^4 (q_0/2)} \right] T^2 e^{-2 \Delta/T} \quad , \quad 1 < h < 3
\quad , \quad T \ll \Delta \,.
\label{eq3}
\eeq
Note that Eq.~(\ref{eq3}) does not require $\Delta$ to be much smaller than $J$.

%%%%%%%%%%%%%%%%%%%%%%%%%%%%%%%%%%%%%%%%%%%%%%%
\subsubsection{Subleading term}
\label{sec:subleading}
%%%%%%%%%%%%%%%%%%%%%%%%%%%%%%%%%%%%%%%%%%%%%%%
We now turn to the term of order $e^{-3 \Delta/T}$. For $1<h<3$ this will be smaller than the $e^{-2 \Delta/T}$ term computed in Sec.~\ref{sec:leading}, while for $3<h<5$ this turns out to be the largest non-zero term.

We will evaluate
\be
c_2(T)=\lim_{\omega\to 0}\frac{ {\rm Im}\big({\cal
    C}_2(\omega)\big)}{\omega}\ .
\ee
This can be cast in the form
\be
c_2(T)=\lim_{\eta\to 0}\left[
  c_2(T,\eta)-{\cal Z}_1c_1(T,\eta)-({\cal Z}_2-{\cal
  Z}_1^2)c_0(T,\eta)\right] ,
\label{c2T}
\ee
where we have defined
\bea
c_n(T,\eta)&=&\lim_{\omega\to 0}{\rm
  Im}\left[\frac{C_{n,n+1}(\omega)+C_{n+1,n}(\omega)}{\omega}\right]\ ,\quad
n=0,1,2\ ,
\eea
and ${\cal Z}_n$ are the contributions of $n$-particle states to the
partition function
\be
{\cal Z}_1=\sum_{p\in \R}e^{-\veps_p/T}\ ,\qquad
{\cal Z}_2=\sum_{p_1 < p_2\in
  \NS}e^{-[\veps_{p_1}+\veps_{p_2}]/T}\ .
\ee
The explicit expressions for $c_n(T,\eta)$ are
\bea
c_0(T,\eta)&=&\sum_q|\langle
q|\sigma^z_0|0\rangle|^2\left[1-e^{-\veps_q/T}\right]
\frac{4\eta\veps_q}{[\veps_q^2+\eta^2]^2}\ ,\nn
c_1(T,\eta)&=&\sum_{p_1<p_2}\sum_q|\langle
p_1,p_2|\sigma^z_0|q\rangle|^2\left[e^{-\veps_q/T}- e^{-(\veps_{p_1}+\veps_{p_2})/T}
  \right]
\frac{4\eta(\varepsilon_{p_1}+\veps_{p_2}-\veps_q)}
     {[(\varepsilon_{p_1}+\veps_{p_2}-\veps_q)^2+\eta^2]^2}\ ,\nn
c_2(T,\eta)&=&\sum_{p_1<p_2<p_3}\sum_{q_1<q_2}|\langle
p_1,p_2,p_3|\sigma^z_0|q_1,q_2\rangle|^2\left[
  e^{-(\veps_{q_1}+\veps_{q_2})/T} - e^{-(\veps_{p_1}+\veps_{p_2}+\veps_{p_3})/T}\right]\nn
&&\qquad\times\
\frac{4\eta(\veps_{p_1}+\veps_{p_2}+\veps_{p_3}-\veps_{q_1}-\veps_{q_2})}
     {[(\veps_{p_1}+\veps_{p_2}+\veps_{p_3}-\veps_{q_1}-\veps_{q_2})^2+\eta^2]^2}, \label{c012}
     \eea
where the form factors are given in Eqn~(\ref{ff}). Note that $c_0 (T, \eta) \rightarrow 0$ as $\eta \rightarrow 0$ because it is not possible to satisfy the energy conservation delta function. Also in this limit, $c_1 (T, \eta) \rightarrow c_1 (T)$ computed in Eq.~(\ref{c1T}). On the other hand $c_1(T)$ vanishes for $3<h<5$ which implies that in this range of magnetic fields no ``disconnected'' contributions arise in Eq.~(\ref{c2T}) and we simply have $c_2 (T) = \lim_{\eta \rightarrow 0} c_2 (T,\eta)$. It is then straightforward to turn sums into integrals and we examine some properties of the resulting expression for $c_2 (T)$ in Appendix~\ref{app:c2}. By contrast, for $1 < h < 3$ $c_2(T,\eta)$ diverges with system size and the disconnected contributions in (\ref{c2T}) need to be taken into account in order to obtain a finite expression. In principle it is possible to obtain a multiple contour integral representation for $c_2(T)$, but here we confine ourselves to a numerical evaluation of the momentum sums. We proceed as follows:
\begin{enumerate}
\item{} We evaluate $c_n(T,\eta)$ for several values of $\eta$ and
system sizes up to $L=256$. We require $\eta$ to be sufficiently large
so that the finite-size corrections are negligible for our largest
system sizes. We find that $\eta\approx 0.1$ is an appropriate order
of magnitude.
\item{} We numerically extrapolate our results to $\eta=0$ using a
third order polynomial.
\end{enumerate}
A useful check on this procedure is obtained by carrying it out for
$c_1(T)$ and comparing it to the numerically evaluated expression
\fr{c1T}, which is the result in the thermodynamic limit at $\eta\to
0$.
\begin{figure}
\centering
\includegraphics[width=0.7\textwidth]{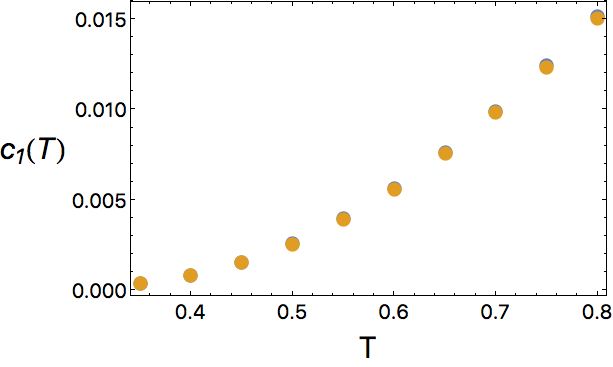}
\caption{$c_1(T)$ for $h=1.5$ and several temperatures. The
  thermodynamic limit result (blue dots) is seen to be in good agreement with the
  extrapolation of numerical results for $L=192$ and $0.1\leq\eta\leq
  0.115$ (yellow dots).}
\end{figure}

\begin{figure}
\centering
\includegraphics[width=0.7\textwidth]{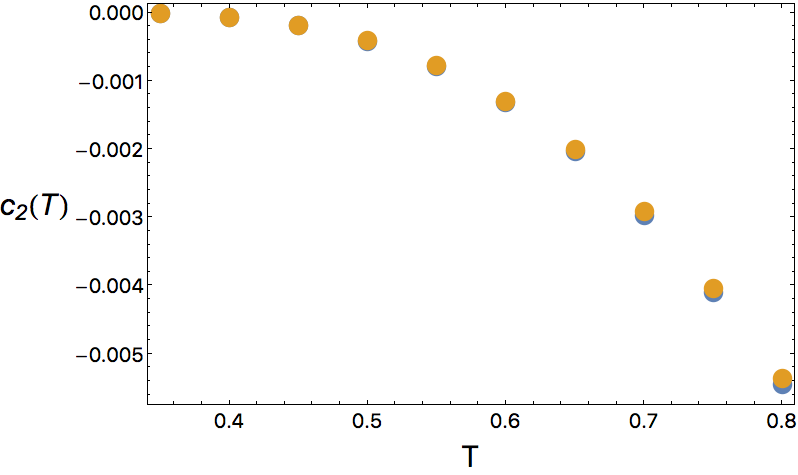}
\caption{$c_2(T)$ for $h=1.5$ and several temperatures. The
results of extrapolating numerical results for $0.1\leq\eta\leq
0.115$ and $L=144$ are in good agreement with those for $L=192$.}
\label{fig:C2332}
\end{figure}

Results for $c_2(T)$ for $h=1.5$ are shown in Fig.~\ref{fig:C2332},
where we compare extrapolations for system sizes $L=144$ and $L=192$.

\section{Quantum criticality}
\label{sec:qc}

In this section we consider the approach to the quantum critical point at $h=1$ in Fig.~\ref{fig:ising}.  It is useful to first review the analysis on the ferromagnetic side, $h<1$, which was presented in Ref.~\onlinecite{Imai14}.
Then as $\Delta \ll J$, the $1/T_1$ rate obeys the universal scaling form
\beq
\frac{1}{T_1} = |a_{hf}|^2 \frac{Z}{T^{3/4}} \Phi_1 (\Delta/T) \quad , \quad h < 1 \quad , \quad T,\Delta \ll J \,, \label{t11}
\eeq
where $Z$ is a non-universal constant, while $\Phi_1$ is a universal function describing the crossovers between the quantum critical and 
renormalized classical regions. For the nearest-neighbor Ising model in Eq.~(\ref{HI}) we take $Z=J^{-1/4}$ for our normalization of $\Phi_1$. While other microscopic
models will have different values of $Z$, the function $\Phi_1 (\Delta/T)$ is independent of the
specific microscopic Hamiltonian.
The limiting forms for $\Phi_1$ in the two regimes are known exactly:
\beq
\Phi_1 (\Delta /T) = \left\{
\begin{array}{ccc} 2.1396\ldots & , & \Delta \ll T \ll J\\
\pi (\Delta/T)^{1/4} e^{\Delta/T} & , & T \ll \Delta \ll J
\end{array}
\right. \,.
\eeq
Furthermore, these theoretical predictions were found to be in good agreement with experimental observations \cite{Imai14}.

Now let us examine the paramagnetic phase $h>1$. As in Eq.~(\ref{t11}), the scaling form is
\beq
\frac{1}{T_1} = |a_{hf}|^2 \frac{Z}{T^{3/4}} \Phi_2 (\Delta/T) \quad , \quad h > 1 \quad , \quad T,\Delta \ll J \,, \label{t12}
\eeq
is expected to describe the crossovers in the NMR relaxation between the quantum disordered and quantum critical regimes.
Matching with Eq.~(\ref{t11}) in the quantum critical regime we have
\beq
\Phi_2 (\Delta/T) = 2.1396 \ldots \quad , \quad \Delta \ll T \ll J\,.
\eeq
For the form of $\Phi_2$ in the quantum disordered regime, we examine only the leading term in Eq.~(\ref{eq3}) in the limit $T \ll \Delta \ll J$.
In this limit, the fermion dispersion in Eq.~(\ref{eq:disp}) takes a relativistic form
\beq
\varepsilon_p = \sqrt{ \Delta^2 + c^2 p^2 }
\label{eq:disp2}
\eeq
with $c= 2 J$. We evaluate the other parameters introduced above Eq.~(\ref{eq3}) for this dispersion and find
\beq
m = \Delta/c^2 \quad, \quad
p_0 = \sqrt{3}\Delta/c \quad , \quad
v = \sqrt{3}c/2\,.
\eeq
Finally, inserting in Eq.~(\ref{eq3}) we obtain
\beq
\frac{1}{T_1} = |a_{hf}|^2 \frac{3 \sqrt{3} T^2}{2 \pi J^{1/4} \Delta^{11/4}} e^{-2 \Delta/T} \quad, \quad T \ll \Delta \ll J,~~ h>1
\eeq
This result is compatible with the scaling form in Eq.~(\ref{t12}), and we have
\beq
\Phi_2 (\Delta/T) = \frac{3 \sqrt{3}}{2 \pi} \left(\frac{T}{\Delta} \right)^{11/4} e^{-2 \Delta/T} \quad , \quad T\ll \Delta \ll J\,.
\label{phi2}
\eeq
Note that the result in Eq.~(\ref{phi2}) applies to a generic ferromagnetic quantum Ising chain near its transverse field quantum critical point, 
and not just the nearest-neighbor model.

\section{Experiments}
\label{sec:exp}

\begin{figure}
\begin{center}
\includegraphics[width=5in,angle=0]{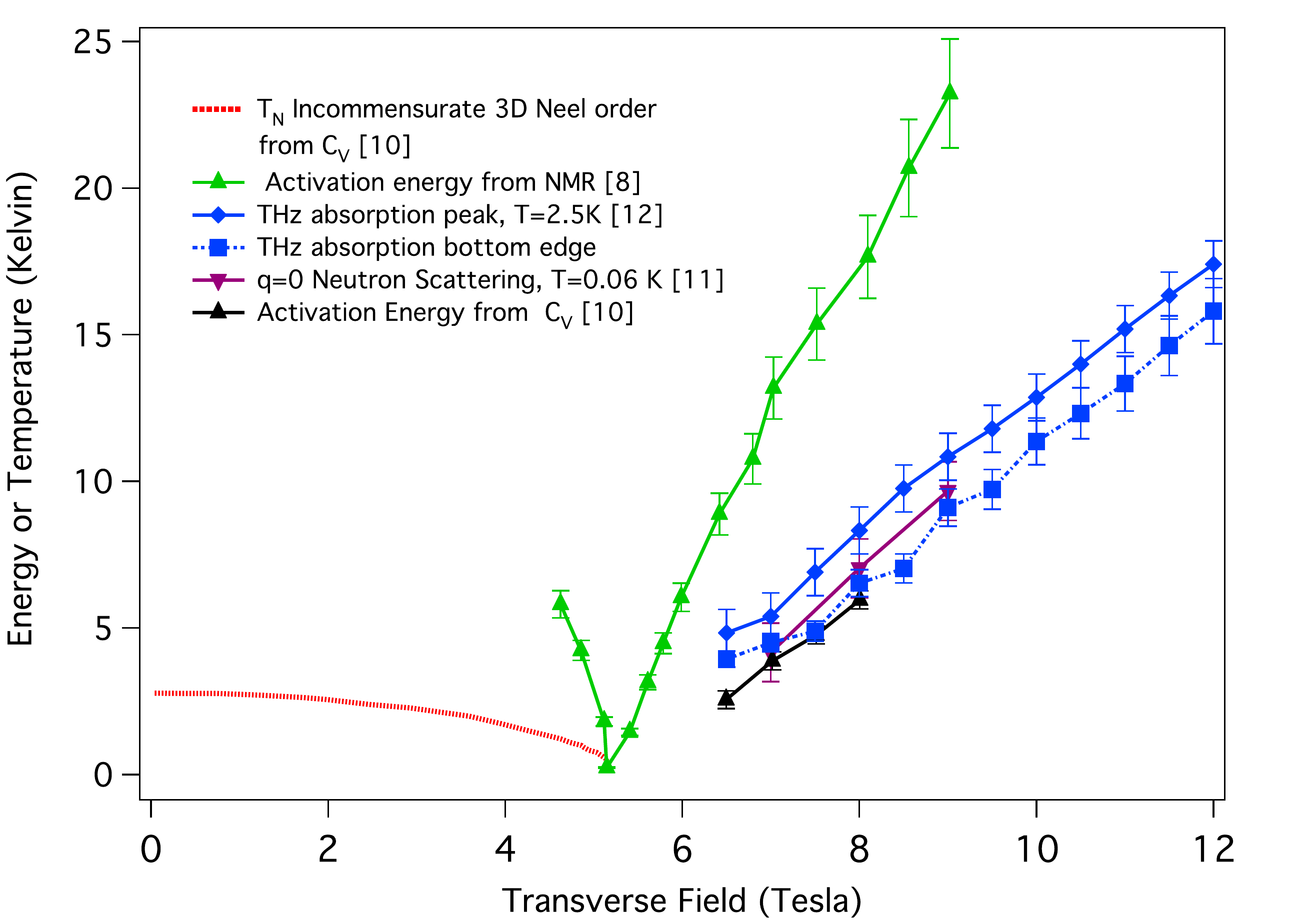}
\end{center}
\caption{Phase diagram and energy scales of the Ising chain system CoNb$_2$O$_6$ as a function of transverse field.   Energy scales from different experimental probes as well as the transition temperature of the 3D incommensurate Neel order are given.   The transition is to a state with ferromagnetic chains that are ordered antiferromagetically in the $b$.  On the paramagnetic side of the transition, one can see an excellent agreement between various spectroscopic probes (THz absorption \cite{Virok2018} and inelastic neutron scattering \cite{robinson2014quasiparticle} and heat capacity  \cite{liang2015heat}).  In contrast one can see clearly that the gap extracted from the temperature dependence of the NMR spin relaxation time is approximately twice as large \cite{Imai14}. } 
\label{fig:Experiments}
\end{figure}
As mentioned above CoNb$_2$O$_6$ has been discovered to be an almost ideal realization of a 1D ferromagnetic Ising chain \cite{Coldea10}.  It is quasi-1D material characterized by zig-zag chains of Co$^{+2}$ ions with effective spin 1/2 moments.  The spins lie in the $ac$ plane at an angle of $\pm31^\circ$ to the $c-$axis \cite{Heid1995,Kobayashi2000} with the chains extending along the $c-$direction.  A dominant ferromagnetic exchange between nearest-neighbor Co$^{+2}$ ions along the $c$ axis cause strong 1D ferromagnetic correlations to develop below $\sim$25 K \cite{Hanawa1994}.   At zero transverse field, weak AF inter-chain exchange interactions stabilize an incommensurate spin-density waves at 2.95 K along the $b$-direction with a temperature-dependent ordering wave vector $\bf{Q}$, and then a commensurate spin-density wave at 1.97 K.  But at temperatures above these scales at zero field (and at temperatures much lower near the 1D quantum critical point (QCP)), the system can be described as a 1D Ising system.   The  effective 1D phase transition to a quantum disordered phase has been inferred to be at the relatively modest critical transverse field of 5.2 T in the $b$ direction.   Note that the 3D phase is believed to extend out slightly past the effective 1D QCP, a feature necessary for the observation of ``kink" bound states in the spectrum near the critical point \cite{Coldea10}.

A number of measurements have been been made of the various energy scales on both sides of the transition of the 1D QCP in CoNb$_2$O$_6$.   We will concentrate on the paramagnetic regime.   As shown in Fig. \ref{fig:Experiments} neutron scattering \cite{robinson2014quasiparticle} and THz absorption \cite{Virok2018} experiments have given evidence for a $q=0$ (or symmetry equivalent) mode which increases in energy roughly linearly with field from the critical point.   This may be identified straightforwardly with the zone center excitation described by Eq. \ref{eq:disp}.   Heat capacity experiments have also been performed \cite{liang2015heat} and data fit to the nearest-neighbor Ising model.  As seen in Fig. \ref{fig:Experiments}, the extracted gap scale from heat capacity is in excellent agreement with the spectroscopic probes.   In contrast to these experiments, the scale of the lowest energy excitation extracted from the temperature dependence of the $1/T_1$ in NMR is greater by approximately a factor of two than the other probes.   As explained above, this data was fit to a activated functional form which was effectively  $1/T_1 \sim \mathrm{exp}(-\Delta_{NMR}/T)$.   However, we have shown here that near the critical point the expectation is in fact  $1/T_1 \sim \mathrm{exp}(-2 \Delta/T)$.   This means that the activation energy  scale from $1/T_1$ will be double that extracted from the other probes.   This is precisely as observed experimentally.   Also note that the differences between energy scales are far bigger than anything that could be explained by inter-chain couplings or 1D vs. 3D regimes of behavior.   The coupling in the transverse $b$ direction has been found to be smaller than 1/60 of $J$ \cite{Cabrera14}.  At low temperature on the paramagnetic side of the transition, this gives a minimum of the dispersion at finite $q_b$, but except very near the critical point this band width in directions perpendicular to the chain is a very small fraction of $\Delta$ \cite{Cabrera14}.  The third direction has frustrated antiferromagnetic couplings and has even smaller effect on the dispersions (although it is presumably responsible for stabilizing different magnetically ordered states at low transverse field \cite{lee2010interplay}).  The scenario put forward in the current work comes with a distinct prediction.  At fields three times the critical field $1/T_1$ should crossover to a form that goes as $ \mathrm{exp}(-3 \Delta/T)$ e.g. a much faster dependence.   This is at fields greater than 15.6 T in CoNb$_2$O$_6$ and should be easily testable.

\section{Conclusions}
\label{sec:conc}

The quantum Ising chain has been an essential model to understand the low frequency, non-zero temperature dynamics of a strongly interacting system \cite{SSAPY97,Essler09,Essler12}. The integrability of the model allows for exact solutions, and yet many local observables exhibit generic dissipative dynamics at long times. Here we have examined the NMR relaxation rates in the quantum disordered region. It is given by the imaginary part of the local spin susceptibility, at frequencies far below the quasiparticle gap, $\Delta$. 
Therefore it is not directly amenable to a quasi-classical computation involving collisions of a dilute gas of quasiparticles \cite{SSAPY97}.
Instead, we showed here that it is dominated by rare processes in which one quasiparticle has sufficient energy to decay into two quasiparticles (and vice versa) near the nucleus. Consequently we found that the NMR relaxation is suppressed by a thermal Boltzmann factor of $\exp(-2 \Delta/T)$ for not too large a transverse field, $1 < h < 3$
(the suppression is stronger for larger $h$). We also computed the precise prefactor of this exponential for the nearest-neighbor Ising chain, and its universal form near the quantum critical point.  Finally we compared our results to the experimental probes.  The scenario put forth here is in excellent agreement with the experimental results.   Spectroscopic and thermodynamic probes show agreement as to the size of the gap, whereas $1/T_1$ from NMR shows an activation energy, which is approximately twice as large.

\subsection*{Acknowledgements}
%********************************************************

This research was supported by the National Science Foundation under Grant No. DMR-1664842 and by the EPSRC under grant EP/N01930X.  Research at Perimeter Institute is supported by the Government of Canada through Industry Canada and by the Province of Ontario through the Ministry of Research and Innovation. SS also acknowledges support from Cenovus Energy at Perimeter Institute. J.S. acknowledges support from the National Science Foundation Graduate Research Fellowship under Grant No. DGE1144152. NPA was supported as part of the Institute for Quantum Matter, an Energy Frontier Research Center funded by the U.S. Department of Energy, Office of Science, Office of Basic Energy Sciences under Award Number DE-SC0019331.  We thank T. Imai, T. Liang, and N.P. Ong for helpful correspondences regarding their data.

\appendix
\section{Evaluation of $e^{-3 \Delta/T}$ contribution}
\label{app:c2}

The expression for $c_2 (T, \eta)$ in Eq.~(\ref{c012}) can be written in the limit $T \rightarrow 0$ as 
\begin{eqnarray}
c_2 (T, \eta) &=&-\frac{2\pi}{2!3!T}\int^{\pi}_{-\pi}\frac{dk_{1}dk_{2}dq_{1}dq_{2}dq_{3}}{(2\pi)^{5}}\vert_{NS}\langle k_{1} k_{2}\vert\sigma^{z}_{0}\vert q_{1}q_{2}q_{3}\rangle_{R}\vert^{2}e^{-(\varepsilon_{q_{1}}+\varepsilon_{q_{2}}+\varepsilon_{q_{3}})/T}
\nn
&&~~~~~~~~~~~~~~~~~~~\times \delta(\varepsilon_{k_{1}}+\varepsilon_{k_{2}}-\varepsilon_{q_{1}}-\varepsilon_{q_{2}}-\varepsilon_{q_{3}})\,, \label{j1}
\end{eqnarray}
where the factorials are combinatoric factors from converting the sums to integrals. From Eq.~(\ref{ff}), the form factor is
\begin{eqnarray}
&& \vert_{NS}\langle k_{1} k_{2}\vert\sigma^{z}_{0}\vert q_{1}q_{2}q_{3}\rangle_{R}\vert^{2}=\frac{(4J^{2}h)^{\frac{1}{2}}\vert 1-h^{2}\vert^{\frac{1}{4}}}{\varepsilon_{k_{1}}\varepsilon_{k_{2}}\varepsilon_{q_{1}}\varepsilon_{q_{2}}\varepsilon_{q_{3}}}
\nn
&&~~~~~~~~~~\times
\left(\frac{\sin\left(\frac{k_{1}-k_{2}}{2}\right)\sin\left(\frac{q_{1}-q_{2}}{2}\right)\sin\left(\frac{q_{1}-q_{3}}{2}\right)\sin\left(\frac{q_{2}-q_{3}}{2}\right)}{\sin\left(\frac{k_{1}-p_{1}}{2}\right)\sin\left(\frac{k_{1}-q_{2}}{2}\right)\sin\left(\frac{k_{1}-q_{3}}{2}\right)\sin\left(\frac{k_{2}-q_{1}}{2}\right)\sin\left(\frac{k_{2}-q_{2}}{2}\right)\sin\left(\frac{k_{2}-q_{3}}{2}\right)}\right)^{2}
\nn
&&~~~~~~~~~~\times
\left(\frac{(\varepsilon_{k_{1}}+\varepsilon_{q_{1}})(\varepsilon_{k_{1}}+\varepsilon_{q_{2}})(\varepsilon_{k_{1}}+\varepsilon_{q_{3}})(\varepsilon_{k_{2}}+\varepsilon_{q_{1}})(\varepsilon_{k_{2}}+\varepsilon_{q_{2}})(\varepsilon_{k_{2}}+\varepsilon_{q_{3}})}{4(\varepsilon_{k_{1}}+\varepsilon_{k_{2}})(\varepsilon_{q_{1}}+\varepsilon_{q_{2}})(\varepsilon_{q_{1}}+\varepsilon_{q_{3}})(\varepsilon_{q_{2}}+\varepsilon_{q_{3}})}\right)^{2} \,. \label{eq:imchi}
\end{eqnarray}

We now attempt to take the $T \rightarrow 0$ limit of Eq.~(\ref{j1}) in a manner similar to the analysis below Eq.~(\ref{c1T}). The integral is dominated by small $q_{1,2,3} \sim \sqrt{T}$. This allows us to make the following approximations
\begin{eqnarray}
\varepsilon_{q_{i}}&\approx& \frac{q^{2}_{i}}{2m}+\Delta \nn
\sin\left(\frac{q_{i}-q_{j}}{2}\right)&\approx& \frac{q_{i}-q_  {j}}{2} \nn
\varepsilon_{q_{1}}+\varepsilon_{q_{2}}&=&2\Delta
\end{eqnarray} 
We can now write Eqn.~(\ref{eq:imchi}) as a product of a term containing the $q_{1}$, $q_{2}$, and $q_{3}$ dependence, with one containing the $k_{1}$ and $k_{2}$ dependence. The $q_{1,2,3}$ integral is sharply peaked 
about momenta $\sim \sqrt{T}$ in the $q_{1}$, $q_{2}$, $q_{3}$ plane, so we can extend the limit of integration over these momenta out to infinity giving us the following
\begin{eqnarray}
&& c_2 (T, \eta)
\approx-\frac{ e^{-3\Delta/T}4\pi J(h^{2}\vert 1-h^{2}\vert)^{\frac{1}{4}}}{3!2!\Delta^{9}2^{16}T}\int^{\pi}_{-\pi}\frac{dk_{1}dk_{2}}{(2\pi)^{2}}\frac{\left((\varepsilon_{k_{1}}+\Delta)^{3}(\varepsilon_{k_{2}}+\Delta)^{3}\right)^{2}}{\varepsilon_{k_{1}}\varepsilon_{k_{2}}(\varepsilon_{k_{1}}+\varepsilon_{k_{2}})^{2}}
\nn
&&~~~~~~~~~~~~~~~~~~~~~~\times
\left(\frac{\sin\left(\frac{k_{1}-k_{2}}{2}\right)}{\sin^{3}\left(\frac{k_{1}}{2}\right)\sin^{3}\left(\frac{k_{2}}{2}\right)}\right)^{2}
\delta(\varepsilon_{k_{1}}+\varepsilon_{k_{2}}-3\Delta)
\nn
&&~~~~~~~~~~~~~~~~~~~~~~\times \int^{\infty}_{-\infty}\frac{dq_{1}dq_{2}dq_{3}}{(2\pi)^{3}}\left(\left(q_{1}-q_{2}\right)\left(q_{1}-q_{3}\right)\left(q_{2}-q_{3}\right)\right)^{2}
e^{-(q^{2}_{1}+q^{2}_{2}+q^{2}_{3})/2mT} \,. \label{eq:firstintegral}
\end{eqnarray}
The $q_{1,2,3}$ integrals evaluate to
\begin{eqnarray}
\int^{\infty}_{-\infty}\frac{dq_{1}dq_{2}dq_{3}}{(2\pi)^{3}}\left(q_{1}-q_{2}\right)^{2}\left(q_{1}-q_{3}\right)^{2}\left(q_{2}-q_{3}\right)^{2}e^{-(q^{2}_{1}+q^{2}_{2}+q^{2}_{3})/2mT}=12(2\pi)^{-\frac{3}{2}}(mT)^{\frac{9}{2}}\,,
\end{eqnarray}
which yields
\begin{eqnarray}
&& c_2 (T, \eta) \approx -\frac{e^{-3\Delta/T} 2J(h^{2}\vert 1-h^{2}\vert)^{\frac{1}{4}}(mT)^{\frac{9}{2}}}{\Delta^{9}(2\pi)^{\frac{1}{2}}2^{16}T}
\nn 
&&~~~\times \int^{\pi}_{-\pi}\frac{dk_{1}dk_{2}}{(2\pi)^{2}}\frac{\left(\left(\varepsilon_{k_{1}}+\Delta\right)^{3}\left(\varepsilon_{k_{2}}+\Delta\right)^{3}\right)^{2}}{\varepsilon_{k_{1}}\varepsilon_{k_{2}}(\varepsilon_{k_{1}}+\varepsilon_{k_{2}})^{2}}
\frac{\sin^{2}\left(\frac{k_{1}-k_{2}}{2}\right)}{\sin^{6}\left(\frac{k_{1}}{2}\right)\sin^{6}\left(\frac{k_{2}}{2}\right)}
\delta(\varepsilon_{k_{1}}+\varepsilon_{k_{2}}-3\Delta)\label{eq:k1k2int}
\end{eqnarray}
Now we have to perform the final integrals over $k_{1,2}$. Because of the singularities in the form factors at small momenta, the integrals have infrared divergencies which need to be treated differently depending upon the value of $h$.

For $3 < h < 5$, the energy conservation delta function in Eq.~(\ref{eq:k1k2int}) prevents 
a divergence. The argument of the delta function does not vanish when either $k_1=0$ or $k_2=0$.
Consequently, the $k_{1,2}$ integrals are finite, and we obtain
\beq
c_2 (T, \eta) \sim T^{7/2} e^{-3 \Delta/T} \quad , \quad T \ll \Delta \quad , \quad 3 < h < 5\,,
\eeq
so that the contribution to $1/T_1$ is  $\sim T^{9/2} e^{-3 \Delta/T}$.

However, for $1 < h < 3$, there are divergences in Eq.~(\ref{eq:k1k2int}). 
The divergences are present when either $k_1 =0$ or $k_2 = 0$. So let us consider the form of Eq.~(\ref{eq:k1k2int}) when both $k_1$ are $k_2$ small. After suitable rescaling of momenta, we obtain an expression of the form
\beq
c_2 (T , \eta) \sim T^{7/2} e^{-3 \Delta/T} \int dk_1 dk_2 \frac{(k_1 - k_2)^2}{k_1^6 k_2^6}
\delta\left( (k_1^2 + 1)^{1/2} + (k_2^2 + 1)^{1/2} - 3 \right)\,.
\eeq
This integral has an effective divergence $\sim \int dk/k^6$. As we discussed below Eq.~(\ref{c012}), this divergence will be cancelled by the other terms in Eq.~(\ref{c2T}). 
A conjectured estimate is obtained by cutting off the divergence at $k \sim \sqrt{T}$, which leads to $c_2 (T) \sim T e^{-3 \Delta/T}$
and therefore a contribution to $1/T_1$ which is  $\sim T^2 e^{-3 \Delta/T}$.

\bibliography{ising.bib}

\end{document}